# Understanding blue shift of the longitudinal surface plasmon resonance during growth of gold nanorods

**Aditya K. Sahu[1], Anwesh Das[1], Anirudha Ghosh[1,2], and Satyabrata Raj[1,*]**

[1] Department of Physical Sciences, Indian Institute of Science Education and Research Kolkata Mohanpur, Nadia 741246, India.

[2] Department of Physics and Astronomy, Uppsala University, P.O. Box 256, SE-751 05 Uppsala, Sweden.

*E-mail: raj@iiserkol.ac.in

**Abstract**

We have investigated in detail the growth dynamics of gold nanorods with various aspect ratios in different surrounding environments. Surprisingly, a blue shift in the temporal evolution of colloidal gold nanorods in aqueous medium has been observed during the growth of nanorods by UV-visible absorption spectroscopy. The longitudinal surface plasmon resonance peak evolves as soon as the nanorods start to grow from spheres, and the system undergoes a blue shift in the absorption spectra. Although a red-shift is expected as a natural phenomenon during the growth process of all nano-systems, our blue shift observation is regarded as a consequence of competition between the parameters of growth solution and actual growth of nanorods. The growth of nanorods contributes to the red-shift which is hidden under the dominating contribution of the growth solution responsible for the observed massive blue shift.

Supplementary material for this article is available.

## 1. Introduction

Metal nanoparticles have generated huge attraction due to their different distinct properties as compared to their bulk counterparts. The metal nanoparticles show outstanding optical properties associated with the tunability of their localized surface plasmon resonance (SPR) in the visible part of the spectrum [1]. SPR of noble metal nanoparticles is the collective oscillation of electrons induced by the electromagnetic radiation of light. The SPR peak wavelength ($\lambda_{max}$) of nanoparticles depends upon the size, shape, composition, and also the dielectric environment around them [2-6]. Many interesting electronic, optical, catalytic properties have cropped up because of surface plasmon oscillations and have led to exciting applications in several branches of science and technology including information storage, optoelectronics, biological imaging, cancer therapy etc. [7-12]. The most challenging task in the synthesis of metal nanoparticles is to keep control over the shape and size of the particles. For example, spherical gold nanoparticles show a strong single absorption band in the visible region at about 520 nm whereas gold nanorods (NRs) show two absorption bands belonging to transverse and longitudinal LSPR [13].

To explain the existing experimental results a few remarkable theoretical works have been proposed previously [2, 14-17] which calculate the extinction spectra (derived from absorbance or transmittance) of metal nanoparticles. From several metal nanoparticles, gold nanorods have drawn significant research interest in the last decade due to their multiple plasmon absorption bands, the transverse surface plasmon resonance (TSPR), and longitudinal surface plasmon resonance (LSPR) [14,17]. The LSPR is highly sensitive to the rod aspect ratio (length to width ratio) and dielectric function of particle and that of the surrounding medium, allowing it to tune across a broad wavelength range [17]. Absorption and scattering phenomena become dominant at different aspect ratios and gold nanorods can be targeted to optical applications depending on their dimensions. In a colloidal solution containing a mixture of spherical and rod-shaped nanocrystals, a UV-Visible absorption measurement will reflect an ensemble average for their contributions in TSPR and LSPR peaks in terms of its width, intensity, and position. The longitudinal peak in gold NRs is much more intense than the transverse peak. Therefore, robust understanding and good synthesis control over nanorod growth are essential for effective material applications.

Synthesis of highly monodispersed metallic nanorods in aqueous solution has been well reported for the last few decades [18-20]. In one of the very initial reports on gold nanorods in aqueous solutions, Yu et al. had shown the necessity of adding a small amount of silver to the growth solution [21]. Removing silver could lead to the formation of longer nanorods (aspect ratio ~ 20) but with low yield ~ 55% [19]. However, after all these years of continuous investigations, the exact growth mechanism of gold nanorods using the colloidal route is elusive, which could be due to the constraint of proper in-situ experimental facilities which do not alter the physicochemical properties of the system. We used silver assisted, seeded growth mechanism using the quaternary ammonium surfactant, cetyltrimethylammonium bromide (CTAB) for the growth of stable monodisperse gold nanorods, which has received the most favorable route as far as cost-effectiveness and mass production. Various groups have adopted different methods to formulate fundamental correlations between aspect ratio, volume, LSPR peak position, FWHM of LSPR peak, etc.[22-24]. This has been achieved by making a fitting of experimental data with theoretical simulations such as Gans approximation [22], Discrete Dipole Approximation (DDA) [23], Finite Difference Time Domain (FDTD) [24] and other numerical approaches.

Here, we report results on the synthesis of monodisperse NRs by seed-mediated growth method. The synthesis process was followed to ensure all the results were reproducible several times.We have discussed our finding of the blue-shift of LSPR and its extinction wavelength and intensity. We have explicitly used non-destructive optical measurement such as UV-visible absorption spectroscopy, and extract all the information by fitting the respective data to understand the correlation between LSPR position and other parameters in the solution [24]. However, the rod solution ensemble itself is a complex system containing different reactants with numerous factors such as environment di-electric properties,[25] physical/chemical interaction at the surface, [26] surface charge, [27] nanorod assembly,[28] interparticle separations,[29] etc. affecting the highly sensitive plasmon peak position. Also, damping at the metal-ligand (capping molecules) interface has broadened the plasmon resonance width [26]. Recent reports have shown that apart from aspect ratio, the net volume of the particle in a solution can substantially alter the extinction spectrum[30,31]. The LSPR can be tuned by adding different additives such as HCl, $Na_2S$ which regulate the growth reaction by changing the chemical equilibrium [32]. Progressive developments in the quest for better understanding the dominant factors have resulted in several reports [17, 22, 24, 30, 33], yet the results are ambiguous and the dynamics of LSPR peak is elusive. As a step forward, here we demonstrate the temporal evolution of LSPR positions and intensity and its correlation with the growth of gold nanorods in the medium.

*2*. **Experimental Methods**

*2.1. Materials*

Tetrachloroauric acid ($HAuCl_4.3H_2O$) (99.9%), L-ascorbic acid (99 %), and Silver Nitrate ($AgNO_3$) (99.99 %) were procured from Sigma-Aldrich. Sodium borohydride ($NaBH_4$) (99.99%) and Cetyl trimethylammonium bromide (CTAB) with purity $\geq$ 98% were procured from Merck for our gold nanorods synthesis. It is important to mention that high purity CTAB as a surfactant is required for the synthesis as it highly influences the anisotropic structure of Au nanorods [34, 35]. Ultrapure distilled water (DI; purification system manufactured by Millipore Corporation) was used for all solutions preparations and experiments.

## 2.2. Synthesis of Gold Nanorods

A two-step seed-mediated process was used to prepare gold nanorods solution as reported in several literatures [18-20]. All solutions were prepared in aqueous solutions, and the entire reaction was carried out in distilled water at ambient conditions. For the preparation of NRs, seed and growth solutions were made as described below.

*Seed Solution.* In a typical synthesis of the Au seed solution, $NaBH_4$ (0.60mL, 0.010M) was added to the solution mixture of CTAB (5mL, 0.20M) and $HAuCl_4$ (5 mL, 0.0005M) under continuous stirring, which resulted in the formation of a brownish-yellow solution. The solution was stirred for a few minutes and kept at 25 °C. The seed solution has been characterized by UV-vis spectroscopy and the size of the seeds has been determined from the Dynamic Light Scattering (DLS) measurement (see supporting information Fig. S1).

*Growth Solution.* A growth solution was prepared by mixing the aqueous solutions of CTAB (5 ml, 0.2M), $HAuCl_4$ (5 ml, 1mM), $AgNO_3$ (50-200μl, 0.0064M), and ascorbic acid (85μl, 0.0788M). The growth solution is characterized by UV-vis spectroscopy [see supporting information Fig. S2 (a)]. On the addition of ascorbic acid to the solution, the growth solution colour changes into colourless. Freshly prepared growth solution has been used for all the measurements as aging of growth solution affect the results. The aging of solutions can affect the colour, intensity and plasmon resonance peak of the nanoparticles [36].

Both the seed and growth solutions have been kept for 10 minutes with magnetic stirrer on before the synthesis of gold nanorods for the proper mixing of the individual components in the solution. To initiate the growth of gold nanorods, the seed solution was added to the growth solution at 25 °C. The colour of the solution changed gradually after a few minutes.

The UV vis spectrum of the mixed solution [see supporting information Fig. S2 (b)] indicates the formation of the rod particles in the mixed solution. The aspect ratio of Au nanorods was controlled by the addition of different amounts of seed solution (12 – 60 µl) and AgNO$_3$ (50 – 200 µl) to the growth solution. Details of synthesis parameters of Au nanorods have been mentioned in supporting information.

*2.3. Characterization*

To determine the extinction spectrum of the gold nanorods in solution and to discuss the temporal evolution of the growth of gold nanostructures, we have measured UV-visible spectra using the HITACHI U-4100 spectrometer. All the spectra have been measured in constant 25 ºC room temperature and care has been taken to maintain the room temperature for all the measurements. A Gaussian distribution function is used to extract the respective data as defined in the following section to study the UV visible spectrum data. The field effect scanning electron microscope (FESEM) image of Au nanorods was obtained with Carl Zeiss Field Emission Scanning Electron Microscope. For imaging, the gold nanorods were extracted from the solution by centrifuging the samples at 5000 rpm for 10 minutes. The solution was re-dispersed in distilled water to remove the excess surfactants and then centrifuged again at 5000 rpm for 10 minutes. Then the surfactants were discarded and the rods were deposited onto ultrasonically cleaned Si (100) substrates using a spin coater. For calculating particles size distribution more than hundred particles were selected from different parts of the images and measured using ImageJ software. For each sample, the sizes of different nanorods from SEM images were measured to obtain the average size and the size distribution of nanorods. Transmission electron microscopic (TEM) images have been collected on JEOL-JEM 2100 F model using a 200 kV electron source.

## 3. Results and Discussion

The UV- visible spectrum of plasmonic nanostructures can indicate the resonant properties of the nanostructures. From the absorbance spectrum, it is possible to derive much information such as full width at half maximum (FWHM) and the shape of the longitudinal SPR band which is in fact excellent measures of size dispersion. For gold nanorods, there are two absorption peaks: longitudinal surface plasmon resonance (LSPR) peak and transverse surface plasmon resonance absorption (TSPR) peak. The structural anisotropy contributes to different degrees of electrons polarization in all directions of the gold nanorods. Electrons generally resonate with light in the range of 510 nm-530 nm along the short axis of the rod, which is called TSPR, and vibration along the long axis of the rod, which varies widely from the visible region to the near-infrared region, named as LSPR. As seen from Figs. 1(a) and (b), TSPR and LSPR peaks begin to emerge with the progression of reaction time. A significant blue shift is observed in LSPR peak against a nearly fixed TSPR peak position.

The TSPR peak has a contribution mostly from the width of the nanorod which varies depending upon the reaction conditions and also from the spherical particles which are present in the final rod solution [19]. When the seed solution is added to the growth solution, there is some amount of excess $NaBH_4$ in the seed solution which gets transferred to the growth solution thereby leading to the formation of additional spherical nanoparticles. With the progression of the reaction, the width of the rod also increases by some amount [19]. These two factors contribute to the position of the TSPR peak and the intensity (the number of individual oscillators) increases with time as the number of particles increases in the solution [22] In the present investigation, the TSPR positions are nearly fixed within a trivial range between 520 to 540 nm. The saturation of the TSPR peak position is also an indication of the stability of the width of the nanorods in the solution.

We have analyzed the growth dynamics of the nanorod solution for seed solution (12 µl) and AgNO$_3$ (100 µl) and derived the information, which is shown in Fig. 2. Fig. 2 (a) shows a typical UV-visible spectrum of gold nanorod with TSPR and LSPR peaks. The longitudinal plasmon peak begins to appear after 3 to 4 minutes of mixing of solution and it has been found that the peak position has blue shifts from 752 nm to 635 nm with an average standard deviation of 0.7 nm as rod growth with time [Fig. 2 (b)]. We can see that LSPR peak position decreases (decrease of wavelength means blue shift) exponentially with time.

Eustis et al.[22] reported that the intensity of the peak is dependent on the net induced dipole moment of a particle. We found that with the progress of the reaction, the length of the rods in solution increases and the increase of length leads to an increase in the number of atoms in the rod. Thus the increase in individual oscillators in the rod leads to an increase in the intensity of LSPR peak initially. However, after reaching a maximum, there is a small decrease in intensity and after which, it becomes stable [Fig. 2 (c)]. The decrease could be due to the fact that during the seeded growth process, the rods first undergo an increase in length, after which they contract by a small value [19] and finally settle down in the most stable configuration depending on the thermodynamics of the environment in which they are growing.

The full width half maximum (FWHM) of the LSPR peak describes the poly-dispersity of the rod solution. The value of FWHM reflects the lifetime of plasmon and from uncertainty relation ($\Delta\omega \times \Delta t \sim 1$); the increase in width of spectra indicates a lesser lifetime of excited plasmon. Furthermore, local field enhancement is dependent on the lifetime of plasmon. Increasing plasmon damping can be seen in increasing FWHM and leads to shorter plasmon lifetime. Fig. 2 (c) shows the FWHM value obtained at a different time of growth of solution confirms the formation of uniform size distribution. We find that peak shapes are symmetric and corresponding FWHM value ranges from 170 to 120 nm with an average standard deviation of 2.5 nm. It has been found that the size distribution becomes narrower and becomes constant after 25 min of growth reaction. The extinction ratio of the longitudinal surface plasmon resonance peak to transverse peak is decreasing with time and becomes stable after 1 hour of growth reaction [Fig. 2 (d)].

Although all the ingredients are essential for Au nanorod growth, the amount of silver nitrate and seed solution are the most effective way to tune the plasmon peaks [18]. We have prepared gold nanorods by adding different amounts of seed and $AgNO_3$ in growth solution while maintaining the concentration of the other precursors constant and performed the experiments to get the properties of samples. In Figure 3, we can see the properties of gold nanorod solution for seed solution (12 µl) and $AgNO_3$ (50,100, and 200 µl). It is important to note that the molarity of $AgNO_3$ is constant (*i.e* 0.0064 M) for all different volumes of $AgNO_3$ added to the seed solution. We have observed the same effect for other volumes of $AgNO_3$, as seen in Fig. 2. Thus, it follows the same characteristics as the previous results. So we can predict the consistency and reproducibility of our results with different parameters. The LSPR Peak positions decrease exponentially with time and increases on increasing the volume of $AgNO_3$ [Fig. 3 (a)]. It has been found that the LSPR peak intensity decreases slightly on the increasing of silver ions [Fig. 3 (b)] whereas the FWHM of the LSPR band is increasing with the silver ion content [Fig. 3 (c)]. The effect of seed solution on LSPR peak wavelength is shown in Fig. 4. The UV visible spectra of the different seed solution for the corresponding $AgNO_3$ volumes are taken after 24 hours of addition of seed solution in to the growth solution. On the addition of more seed solution, the LSPR peak wavelength increases but follows an exponential path with time as mentioned earlier (see supporting information). Based on our observation, we have tuned the LSPR peak wavelength from 600-820 nm by changing the volume of seed solution and $AgNO_3$ solution which effectively change the concentration of reagent per unit volume of the mixed solution in the synthesis process of gold nanorods. We can also tune the LSPR peak wavelength from 600 to 900 nm by changing the concentration of the synthesis parameters [12, 18, 21].

It has been found that during the growth of Au nanorods, LSPR peak position shifted from 780 nm to 650 nm over the period of 60 minutes (reaction time). It undergoes a constant blue shift as observed in Fig. 1. This behavior is inconsistent with the basic understanding of the nanorod growth mechanism. As the rods grow from spheres, the LSPR peak position should, in fact, show a red-shift rather than blue-shift (when we assume other parameters remain unchanged) [17]. The reason for such behavior is that with the increase of the length of the nanorod, the confinement of electrons along the length (long axis) of the rod decreases. As a result, the longitudinal peak should show red-shifting behavior [37]. However, in our case, LSPR peak position undergoes continuous blue-shift. Our results can be compared with that of Sau et al.,[19] where they have shown the behavior of LSPR position based on aspect ratio only (derived from electron microscopy analysis). However, Recio et al.[38] considered the effect of the surrounding medium and showed that in non-dynamical scenarios, the blue-shift of LSPR position could be explained by taking into account the interaction between the collective longitudinal oscillation of electrons and the surrounding medium. Various investigations [39, 40] revealed that the surfactants could undergo reorganization depending on the salt concentration and thermodynamic environment in which they are growing. This reorganization can also lead to the variation in the dielectric constant of the medium. In other way around, theoretical simulations by Link et al.[17] also revealed the impeccable contribution of the dielectric medium in the shift of LSPR peak. Hence, it is clear from these reports that a decrease in aspect ratio is not enough to explain the blue shift unlike that showed by Sau et al.[19] and we need to consider the surrounding medium as well. To understand the role of the medium (interaction between medium and longitudinal oscillation of electrons as reported by Recio et al.[38]) experimentally, we tuned the concentration of monomers in the solution by diluting growth solution with distilled water while keeping the total amount of gold monomers fixed.

Various amount of distilled water (x ml) has been added to the growth solution used for the synthesis of gold nanorods to understand the effect of the surrounding. Then seed solution was added to the respective diluted growth solution to initiate the growth process. The sample B1 was prepared by adding 60 µl of seed solution into a growth solution consisting of aqueous solutions of CTAB (5 ml, 0.2M), $HAuCl_4$ (5 ml, 1 mM), $AgNO_3$ (200 µl, 0.0064 M), ascorbic acid (85 µl, 0.0788 M) using the same synthesis method as described earlier (i.e. x = 0 ml). Here we prepared 20 other samples using identical amounts of monomer but with increasing x (i.e. the volume of $H_2O$) in growth solution from 2.5 ml in B2 to 50 ml in B21 (see supporting information for detail $H_2O$ concentration of B1 – B21 samples). All these samples were obtained after approximately 1 hour from the addition of seed solution to the growth solution and characterized by UV-visible-NIR spectroscopy and electron microscopy.

It is clearly observed from Fig. 5 (a) that with progressive dilution with $H_2O$ of the reacting solution, each LSPR peak undergoes a blue shift as compared to its previous concentrated sample. The peak positions of all the samples have been fitted and found to obey exponential decay behavior, as shown in Fig. 5 (b). It is worthwhile to mention that the point of addition of water is crucial in observing a wide shift in LSPR peak position (from ~ 790 nm to ~ 690 nm, as shown in Fig. 5). When the solution was diluted after the formation of rods, the LSPR peaks did not show any shift; the only observable feature was a decrease in the peak intensity due to a decrease of nanorods per unit volume of solution.

Fig. 6 shows the scanning electron microscope (SEM) images of six intermediate samples between B1 and B21. The histogram of the particle size distribution and the average particle size has been calculated from the Gaussian function fitting of the distributions (as shown in the supporting information). The transmission electron microscope (TEM) images of few intermediate samples are shown in Fig. 7. We have plotted the size distribution of the rods based on the TEM images (see supporting information). It has been found that the aspect ratios of the rods are similar in both the TEM and FESEM images, whereas the rods length and width measured with FESEM is bigger than TEM due to lower resolution in SEM measurements. This difference arises from a combination of inconsistent calibration of magnification, and the different ways contrast is produced in the microscopes [41]. The high-resolution image of a single Au nanorod is also shown in inset of each TEM image. The energy dispersive x-ray spectroscopy (EDS) analysis of few samples are also shown in supporting information (Fig. S7). The analysis of rod sizes from SEM images are shown in Fig. 8. The error bars signify the standard deviation in size distribution calculated for approximately 100-rod particles, selected from different parts of the image. Since the rods grow from its initial stage of spheres in the solution, the addition of seed solution to the growth solution initiate the formation of the rod particles. There are 5-10% of sphere particles present in the solution. We have separated the gold nanorods from the synthetic solution by centrifuging the samples and discarded the excess surfactant [42]. The data points for the variation in length do show an overall decreasing trend, whereas the trend is reversed in the width of the particles; however they are not monotonous in nature. The variation in width does not indicate a proper increasing pattern. For all particles, the width values ranges inside or outside the standard deviation shift around 30 nm. In the growth process, we can also note that the TSPR location does not change. The gold nanorods of B5, B9, and B21 have the same diameter of 30 nm, while they show different 734.4 nm, 697.3 nm, and 689.3 nm plasmon resonances. Thus by taking the width into account, we cannot predict the blue shift of LSPR peak position. There is a blue shift of 100 nm when the aspect ratio value changes by 0.28, which is within the standard deviation limit. We can also observe that for B5 and B9 almost the same aspect ratio shows plasmon peaks of 734.4 nm and 697.3 nm, which is blue shifted despite the same aspect ratio. Thus variation of aspect ratio approach does not perfectly replicate the entire growth process corresponding to the LSPR. For all particles, those fall within or outside the quasi-static limit; the aspect ratio always shows a linear relationship with LSPR maximum position [17,30]. Recent works by various groups [30,33] also show the strong dependence of particle volume on the optical extinction spectrum. It has been further established that the formula for relating the various parameters of the rods and spectral position of the LSPR can be done on the basis of aspect ratio, the dielectric constant of surrounding environment and volume in a combination thereof [17,22,24]. Interestingly, the linear curve fittings do not reveal any significant variation in dimension and aspect ratio although the blue shift is quite substantial (~ 100 nm variation between B1 and B21 samples). This indicates that the LSPR peak is no longer explicitly dependent on the rod size; rather the medium plays an important role in tuning the LSPR peak. Based on various experimental and theoretical interpretations [17,19,22,25,26,43,44] our analysis suggests that the blue shift trend of LSPR peak in the present investigation does indeed depend on the temporal evolution of dielectric constant of the surrounding medium. Our results can be compared to those obtained by Near et al.[30] where they have reported that additional parameters are needed to correlate the LSPR position and aspect ratio.

## 4. Conclusion

We have synthesized and grown nanorods from spherical gold nanoparticle seeds. We observed characterized two plasmon peaks in the absorbance spectra of gold nanorods. The first peak observed near ~ 520 nm corresponds to the transverse surface plasmon resonance, which corresponds to the short axis of the ellipsoid of nanorods. It stays around ~ 520 nm for all growth conditions. The second peak was observed in the range of ~ 550 nm and above on the wavelength scale in the near-infrared region. This is the longitudinal surface plasmon resonance, localized to the long axis of the ellipsoid of nanorods. We observed that the longitudinal plasmon peak is blue-shifted during its growth process. The blue shift is not due to anisotropic or tailored nanorods size and shape, but this can be explained by the change of dielectric permittivity of the surrounding nanorods media. To understand it clearly, we carried out water dilution experiments and found that the blue shift is imminent even without any substantial change in rod size and mostly depends on the physicochemical properties of the surrounding medium.

**Acknowledgements**

The author, AKS acknowledges CSIR, Government of India for financial support.

**Data availability**

The data that support the findings of this study are available upon request from the authors.

**ORCID iDs**

Aditya K. Sahu: https://orcid.org/0000-0003-2405-5906

Satyabrata Raj: https://orcid.org/0000-0003-2193-3152

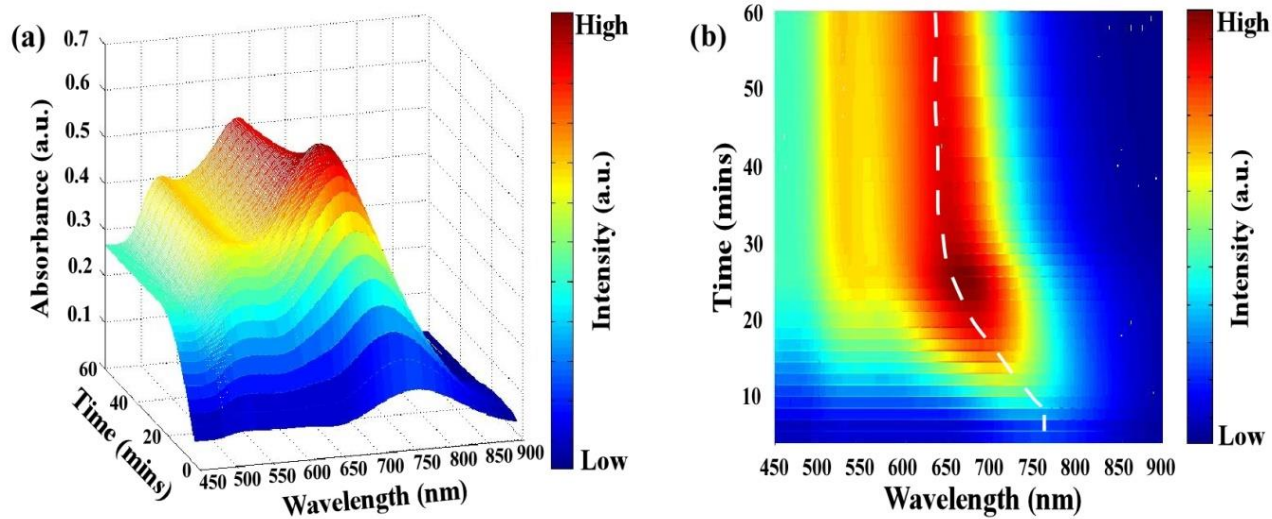

**Figure 1.** (a) Time-dependent absorption curves taken by UV-Vis NIR spectrophotometer of gold nanorods of seed solution (12 μl) and AgNO$_3$ (100 μl) while growth is on the process. (b) Top view of the variation of LSPR peak position w.r.t growth reaction time. The dashed line is the guide to an eye for change of the LSPR peak position.

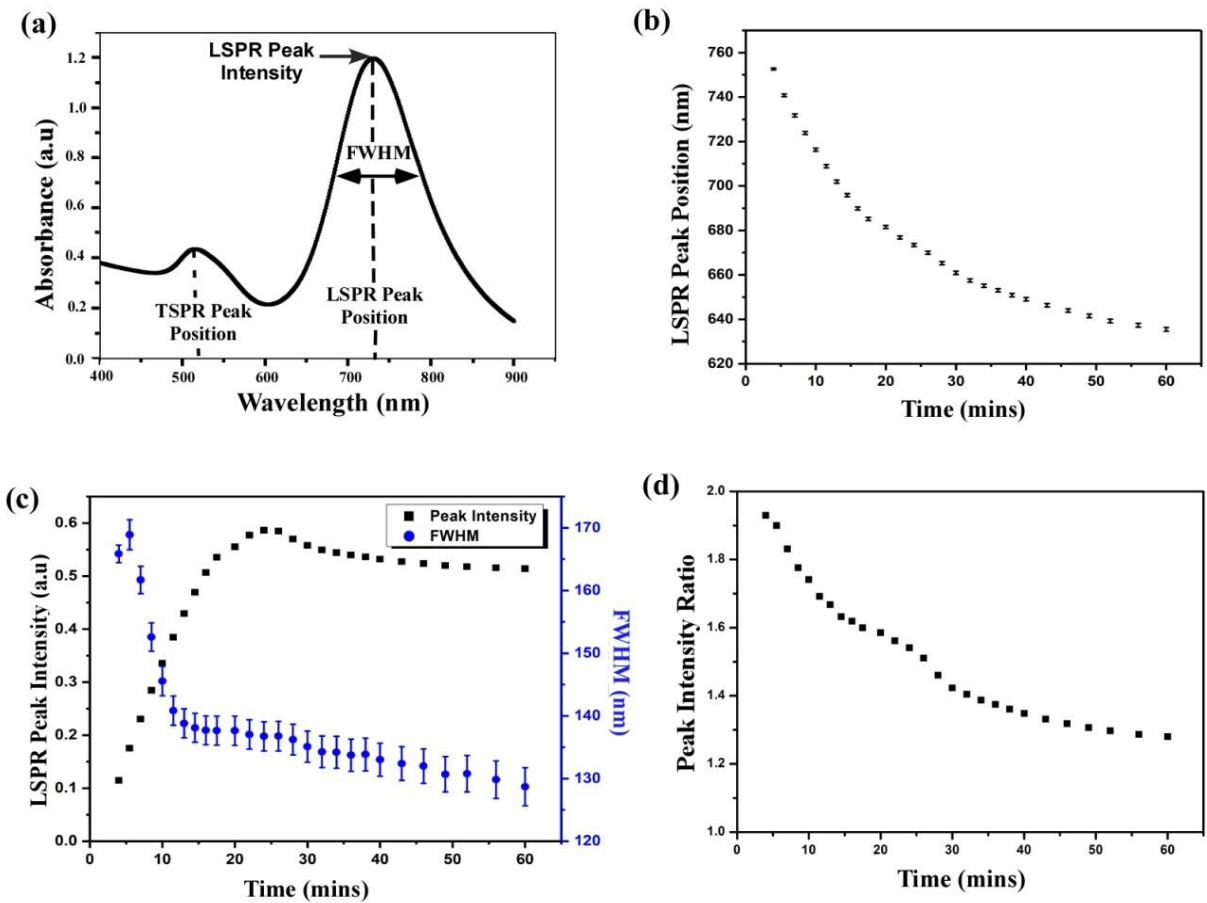

**Figure 2.** (a) Typical UV visible spectrum of a Gold nanorod solution. (b) Variation of longitudinal plasmon band maximum with growth time. (c) Time development of the LSPR Peak intensity (black) and the FWHM (blue) (d) Relative peak intensity of LSPR and TSPR against reaction time of gold nanorods of seed solution (12 μl) and $AgNO_3$ (100 μl).

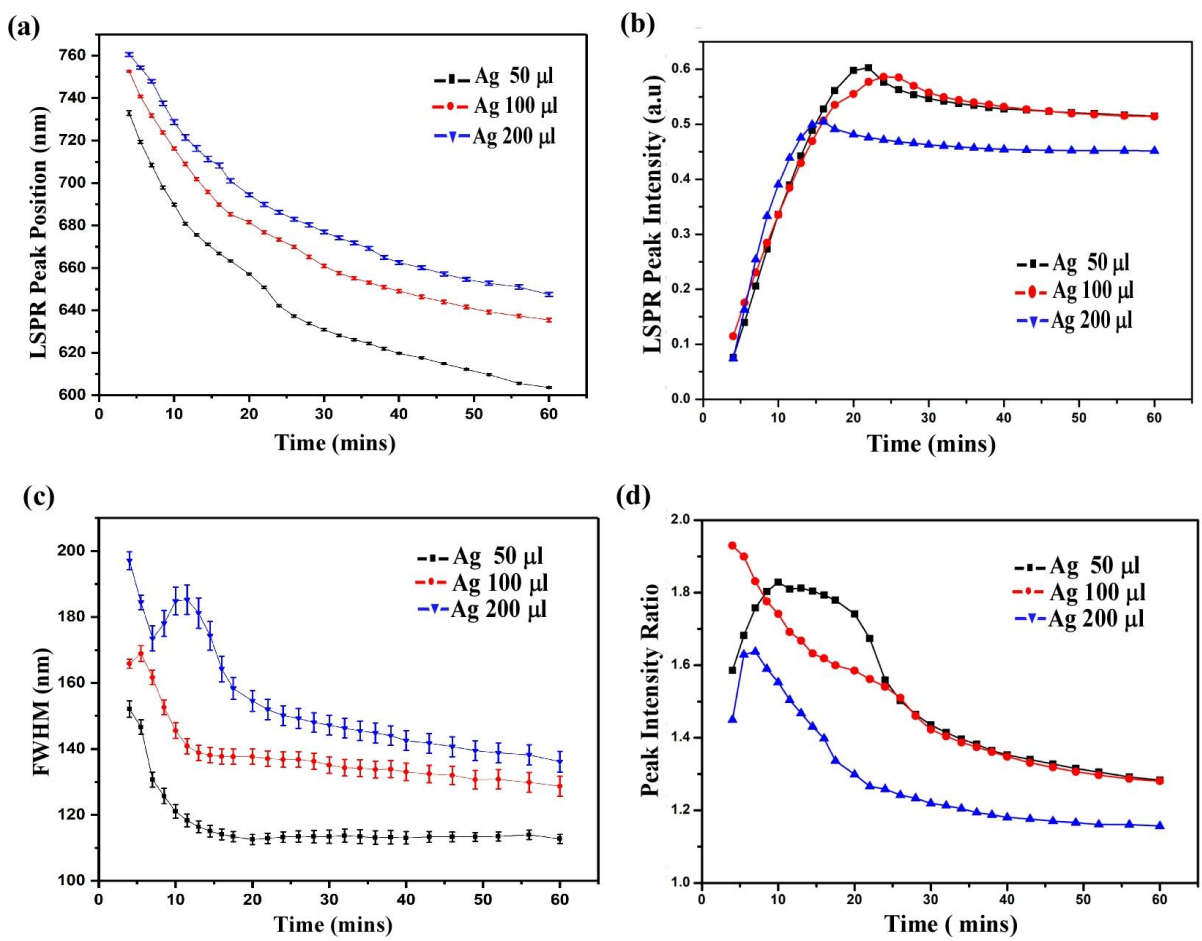

**Figure 3.** Different data analysis parameters vs. time for 12 μl seed concentration with different AgNO$_3$ concentration. (a) Variation of LSPR Peak position with time, (b) variation of LSPR Peak intensity with time, (c) FWHM of LSPR peak with time, and (d) peak intensity ratio with time.

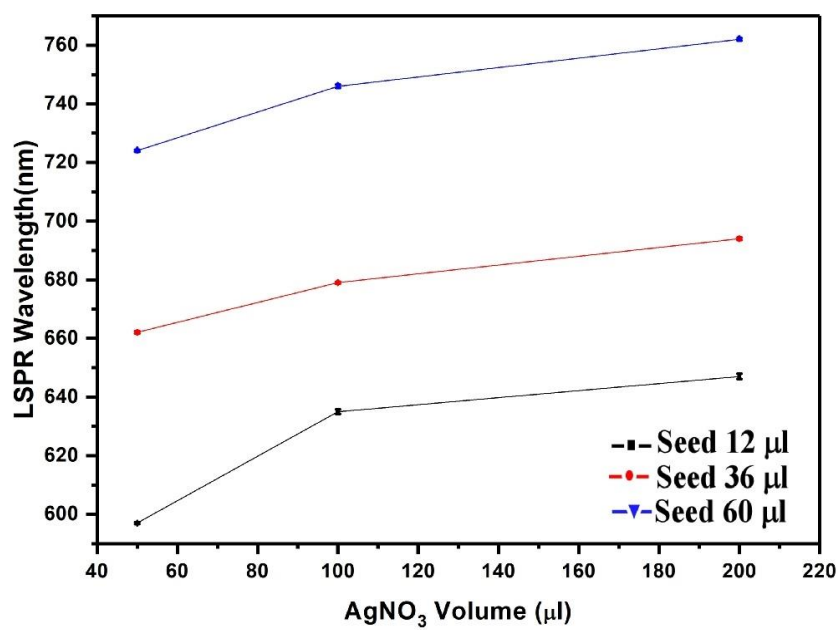

**Figure 4.** Effect of Seed and AgNO₃ volume concentration on LSPR wavelength.

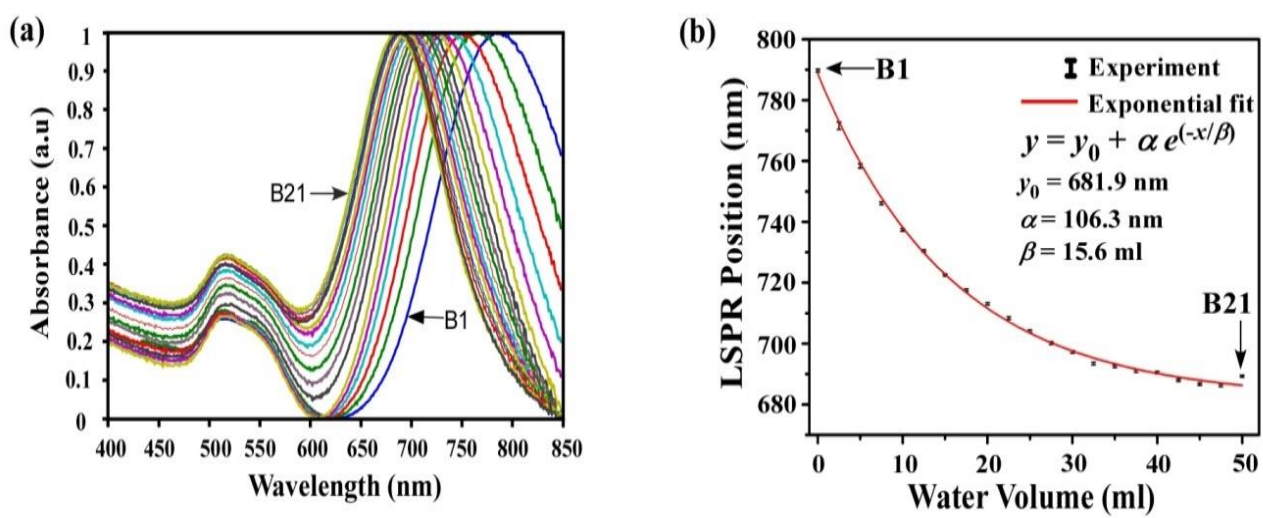

**Figure 5.** (a) UV Visible spectra of 21 nanorod samples (B1 to B21), following adding distilled water to the growth solution in steps of 2.5 ml. B1 dilutes with 0 ml of distilled water, B2 (2.5 ml), B3 (5.0 ml), B4 (7.5 ml), B5 (10 ml), .... B9 (20 ml), B13 (30 ml), B17 (40 ml) and B21 (30 ml). (b) Variation of LSPR peak position with $H_2O$ dilution.

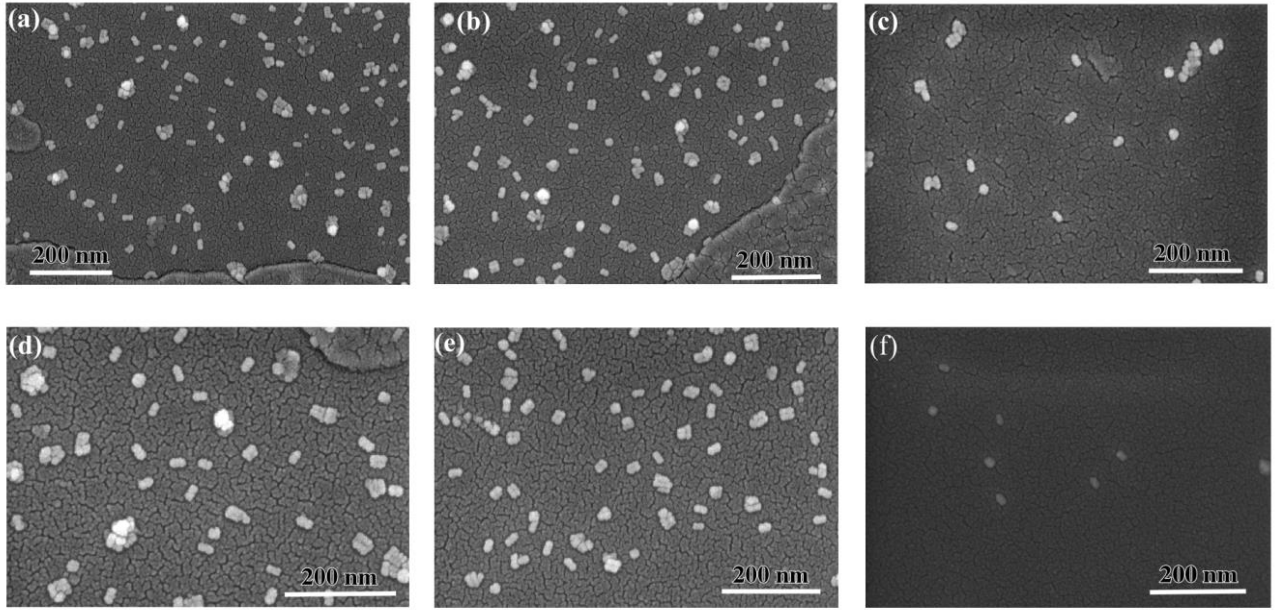

**Figure 6.** FESEM images of (a) B1, (b) B5, (c) B9, (d) B13, (e) B17, and (f) B21 sample.

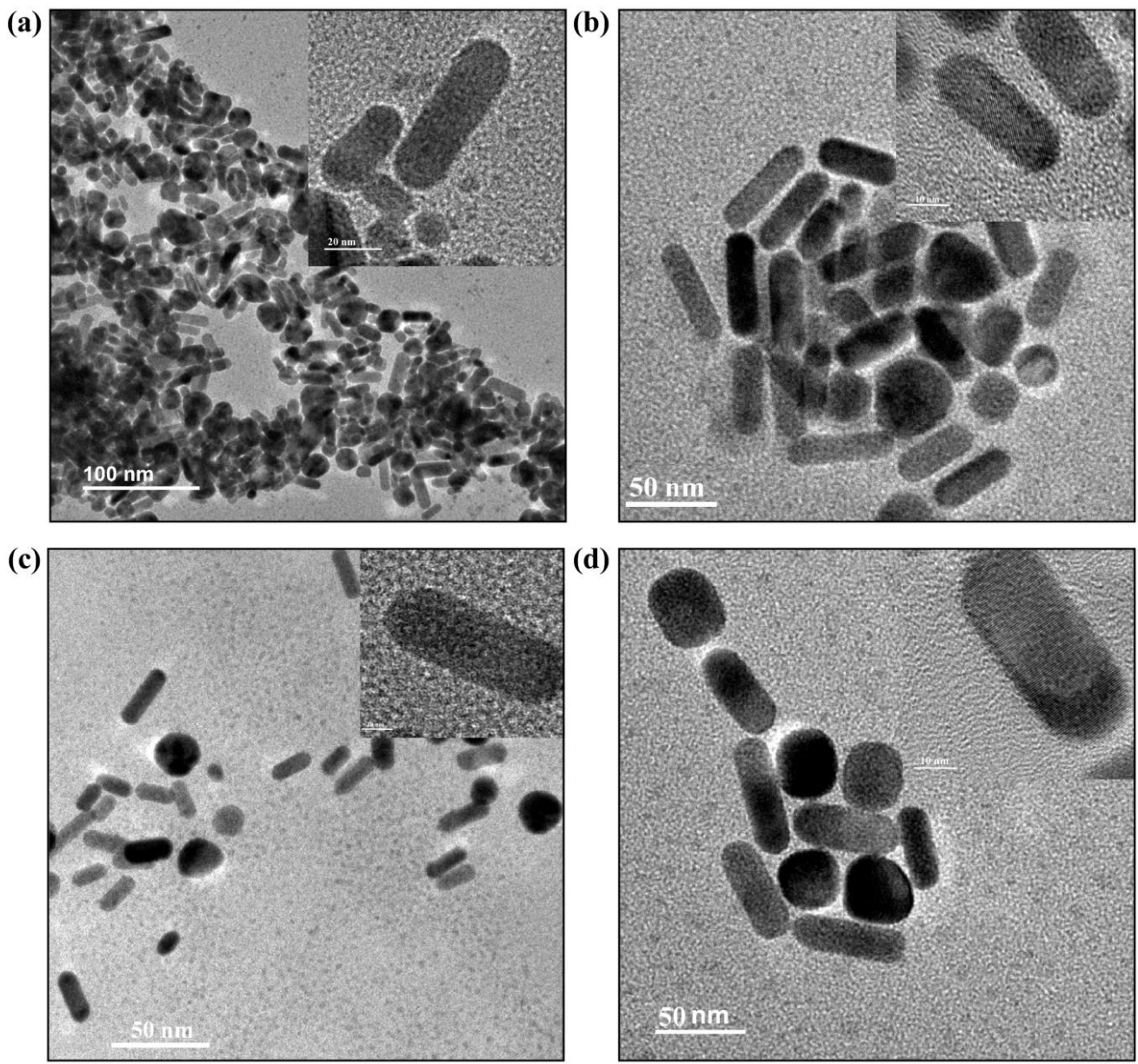

**Figure 7.** TEM images of (a) B1, (b) B5, (c) B9, and (d) B13 samples. Inset of each TEM image shows the high-resolution image of a single Au nanorod.

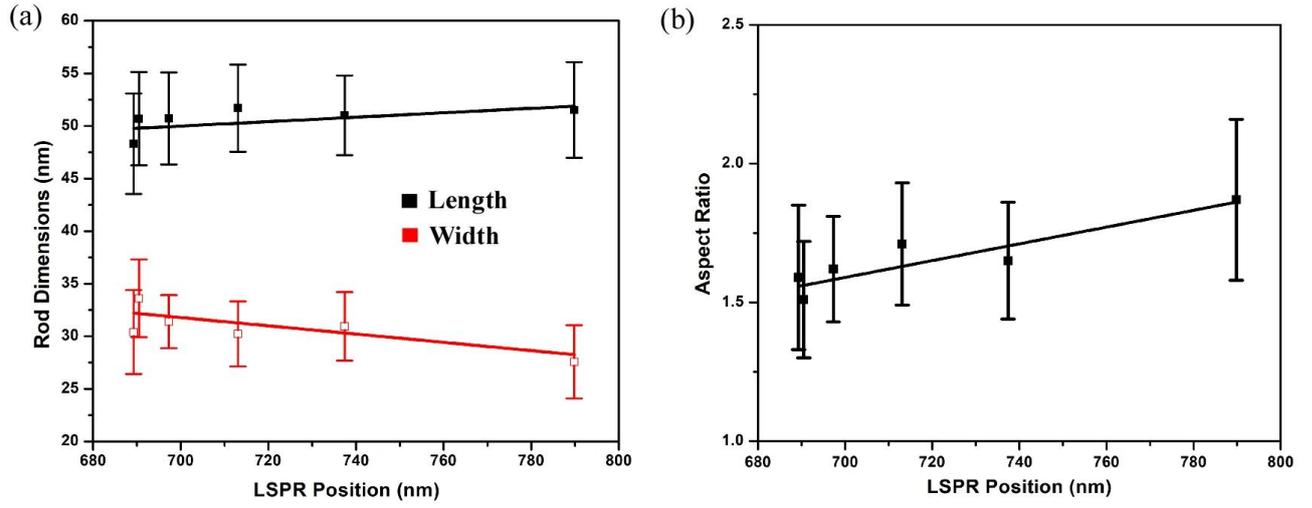

**Figure 8.** (a) Plots of LSPR peak position vs nanorod dimensions (b) LSPR peak position vs aspect ratio of nanorods of B1, B5, B9, B13, B17, and B21 samples.

**Supporting Information**

# Understanding blue shift of the longitudinal surface plasmon resonance during growth of gold nanorods

*Aditya K. Sahu[1], Anwesh Das[1], Anirudha Ghosh[1,2], and Satyabrata Raj[1,]\**

[1]Department of Physical Sciences, Indian Institute of Science Education and Research Kolkata, Mohanpur, Nadia 741246, India

[2]Department of Physics and Astronomy, Uppsala University, P.O. Box 256, SE-751 05 Uppsala, Sweden.

\* Corresponding Author E-mail: raj@iiserkol.ac.in

**UV-Vis spectroscopy of seed, growth, and mixed rod solutions**

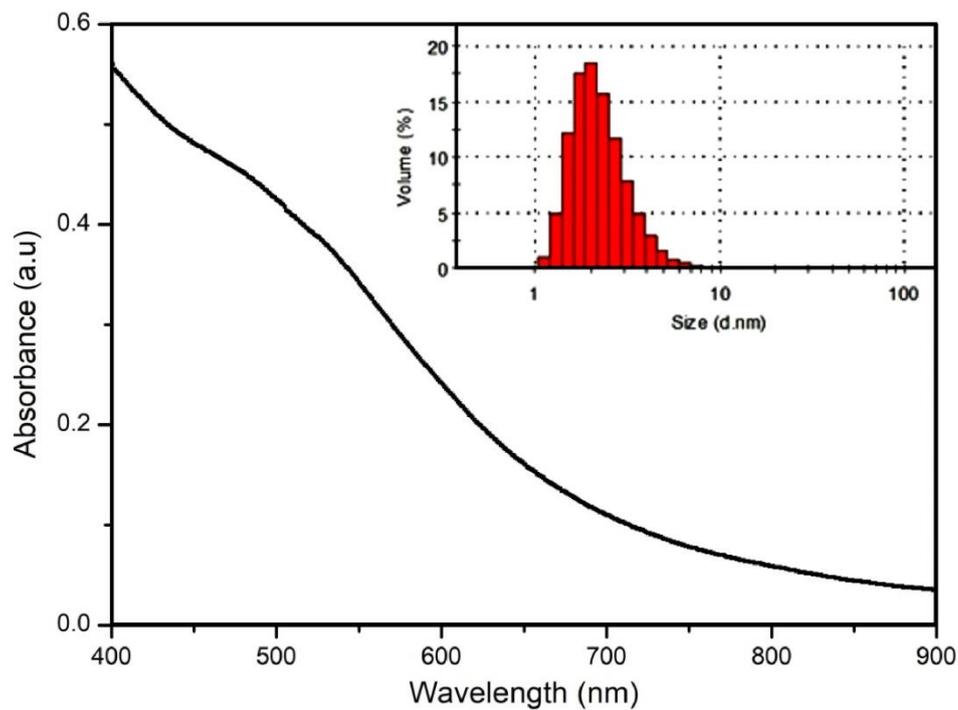

**Figure S1** UV-vis spectroscopy of seed solution. Inset shows the DLS measurement of seed solution.

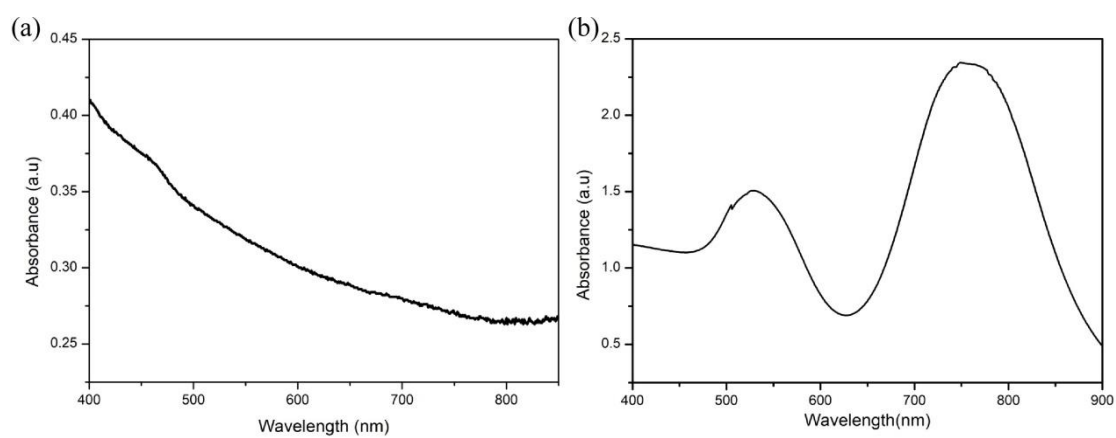

**Figure S2** UV-vis spectroscopy of (a) growth solution and (b) mixed solution (after adding both seed and growth solutions) indicating formation of rods in the mixed solution.

## Synthesis Parameters of Gold Nanorods

A growth solution was prepared by mixing the aqueous solutions of CTAB (5 ml, 0.2 M), $HAuCl_4$ (5 ml, 1 mM), $AgNO_3$ (50-200 μl, 0.0064 M), and ascorbic acid (85 μl, 0.0788 M). We used the $AgNO_3$ volume of 50 μl, 100 μl, and 200 μl in the growth solution. The final volume of the growth solution changed accordingly when different amount of the $AgNO_3$ as mentioned above added to the growth solution while keeping all other parameters same. The Au seed solution was prepared by adding $NaBH_4$ (0.60 mL, 0.010 M) to the solution mixture of CTAB (5 mL, 0.20 M) and $HAuCl_4$ (5 mL, 0.0005 M). We used different seed solution of 12 μl, 36 μl, and 60 μl to mix with the growth solution for the preaparation of gold nanorods. For example, we have taken 36 μl of seed solution and different volumes of $AgNO_3$ to get different nanorods. Other parameters have been kept the same for all the synthesis conditions.

**Table - S1**

Synthesis parameters of nanorods for 36 μl seed solution

|  | Seed solution | Growth solution | | | |
|---|---|---|---|---|---|
| Sample No. | Volume | CTAB | $HAuCl_4 \cdot 3H_2O$ | Ascorbic acid | $AgNO_3$ |
| 1 | 36 μl | 0.2 M, 5 ml | 0.001 M, 5 ml | 0.0788 M, 85 μl | 0.0064 M, **50 μl** |
| 2 | 36 μl | 0.2 M, 5 ml | 0.001 M, 5 ml | 0.0788 M, 85 μl | 0.0064 M, **100 μl** |
| 3 | 36 μl | 0.2 M, 5 ml | 0.001 M, 5 ml | 0.0788 M, 85 μl | 0.0064 M, **200 μl** |

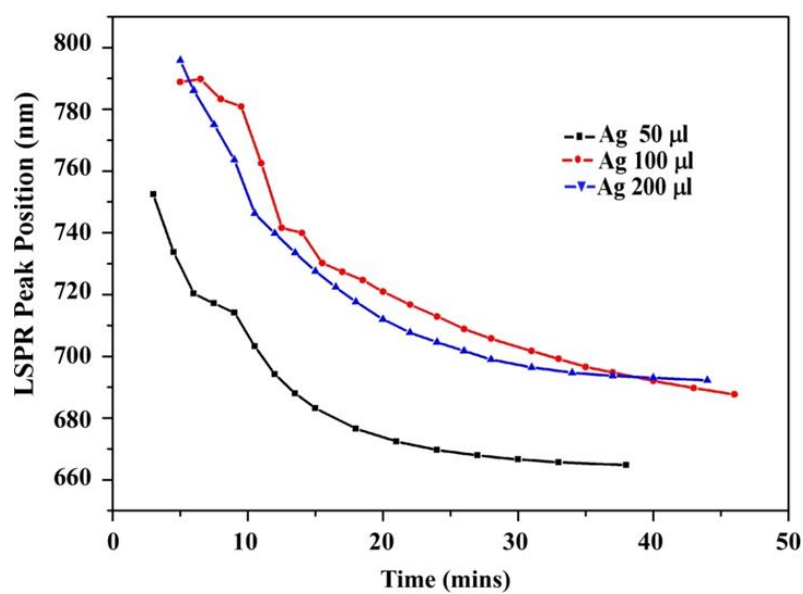

**Figure S3** Variation of LSPR Peak position with time for seed volume 36 μl with different volume of $AgNO_3$

We have taken 60 µl seed solutions and different volumes of AgNO$_3$ to get different nanorods. Other parameters have been kept the same for all the synthesis conditions.

**Table S2**

Synthesis parameters of nanorods for 60 µl seed solution

| Sample No. | Seed solution Volume | Growth solution | | | |
|---|---|---|---|---|---|
| | | CTAB | HAuCl$_4$.3H$_2$O | Ascorbic acid | AgNO$_3$ |
| 1 | 60 µl | 0.2 M, 5 ml | 0.001 M, 5 ml | 0.0788 M, 85 µl | 0.0064 M, **50 µl** |
| 2 | 60 µl | 0.2 M, 5 ml | 0.001 M, 5 ml | 0.0788 M, 85 µl | 0.0064 M, **100 µl** |
| 3 | 60 µl | 0.2 M, 5 ml | 0.001 M, 5 ml | 0.0788 M, 85 µl | 0.0064 M, **200 µl** |

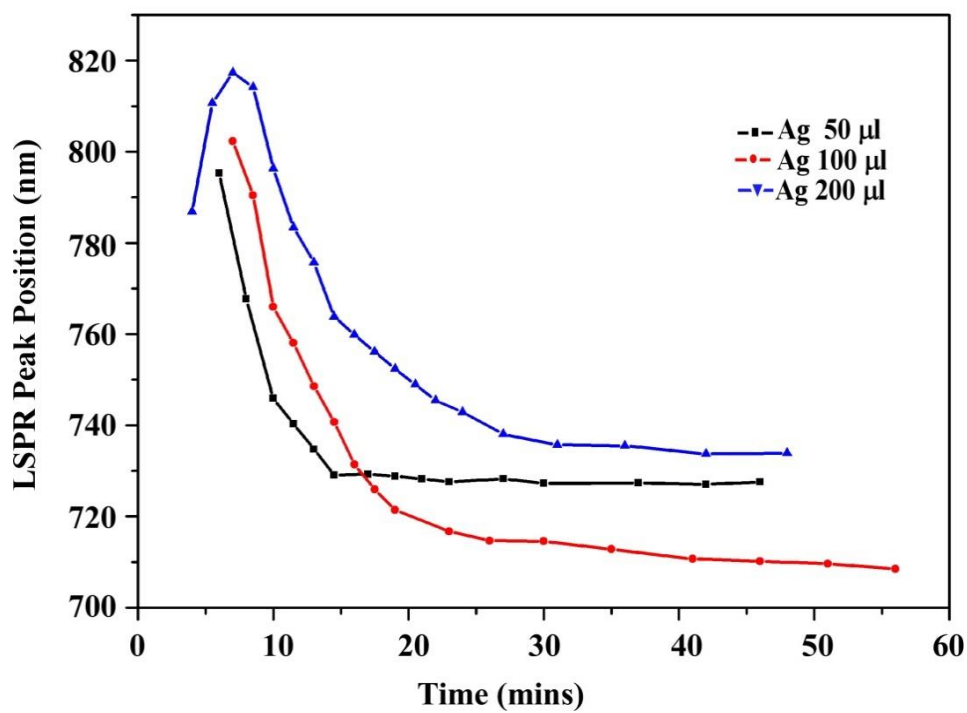

**Figure S4** Variation of LSPR Peak position with time for seed volume 60 µl with different volume of AgNO$_3$.

**Table S3**

The amount of DI water added to the growth solution to control the LSPR peak position. The dimensional parameters corresponding to LSPR position are also mentioned for the nanorods.

| Sl. No. | Sample | x (mL) | LSPR max (nm) | Length (nm) | Width (nm) | Aspect ratio |
|---|---|---|---|---|---|---|
| 1 | B1 | 0 | 789.8 | 51.52 ± 4.55 # | 27.57 ± 3.48 # | 1.87 ± 0.29 # |
| | | | | 50.28 ± 5.43 * | 26.93 ± 4.35 * | 1.87 ± 0.37 * |
| 2 | B2 | 2.5 | 771.6 | - | - | - |
| 3 | B3 | 5.0 | 758.4 | - | - | - |
| 4 | B4 | 7.5 | 746.1 | - | - | - |
| 5 | B5 | 10 | 737.4 | 51.00 ± 3.78 # | 30.94 ± 3.26 # | 1.65 ± 0.21 # |
| | | | | 49.48 ± 4.65 * | 27.87 ± 5.27 * | 1.77 ± 0.37 * |
| 6 | B6 | 12.5 | 730.5 | - | - | - |
| 7 | B7 | 15 | 722.5 | - | - | - |
| 8 | B8 | 17.5 | 717.6 | - | - | - |
| 9 | B9 | 20 | 713.1 | 51.69 ± 4.15 # | 30.23 ± 3.09 # | 1.71 ± 0.22 # |
| | | | | 49.96 ± 5.73 * | 28.07 ± 5.76 * | 1.78 ± 0.42 * |
| 10 | B10 | 22.5 | 708.4 | - | - | - |
| 11 | B11 | 25 | 704.0 | - | - | - |
| 12 | B12 | 27.5 | 700.1 | - | - | - |
| 13 | B13 | 30 | 697.3 | 50.71 ± 4.37 # | 31.40 ± 2.53 # | 1.62 ± 0.19 # |
| | | | | 49.82 ± 4.74 * | 29.52 ± 5.03 * | 1.69 ± 0.33 * |
| 14 | B14 | 32.5 | 693.5 | - | - | - |
| 15 | B15 | 35 | 692.6 | - | - | - |
| 16 | B16 | 37.5 | 691 | - | - | - |
| 17 | B17 | 40 | 690.5 | 50.68 ± 4.45 | 33.61 ± 3.69 | 1.51 ± 0.21 |
| 18 | B18 | 42.5 | 688 | - | - | - |
| 19 | B19 | 45 | 686.7 | - | - | - |
| 20 | B20 | 47.5 | 686.2 | - | - | - |
| 21 | B21 | 50 | 689.3 | 48.31 ± 4.78 | 30.40 ± 3.99 | 1.59 ± 0.26 |

*Data from (#) FESEM and (*) TEM images of the sample*

**Length Distribution of Gold Nanorods on the basis of FESEM Image**

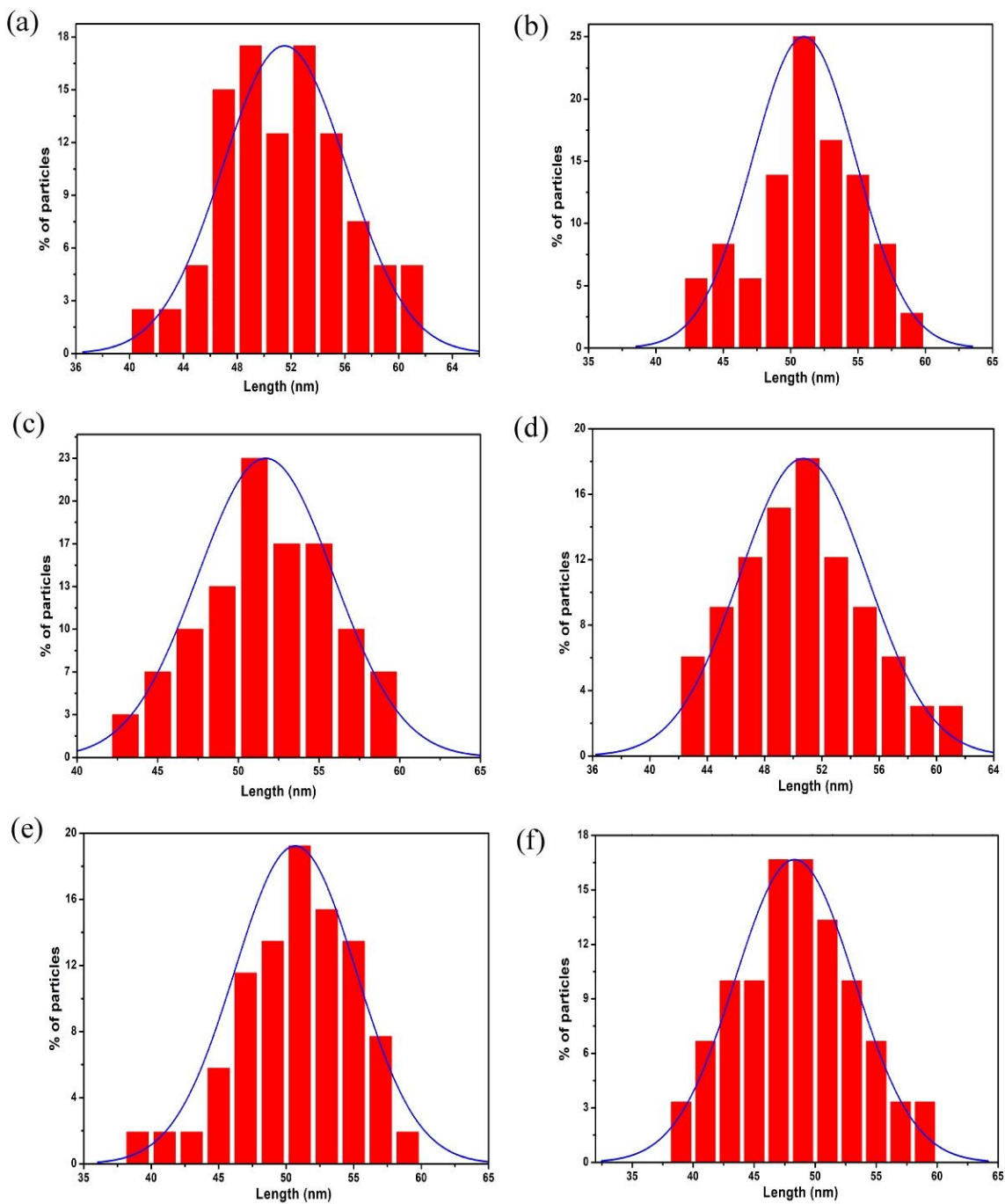

**Figure S5**. The histogram shows the length distribution of gold nanorods. The average particle size has been calculated from the Gaussian distribution function fitting as shown by the solid blue curve for the FESEM images (Fig. 6 of manuscript) of sample (a) B1, (b) B5, (c) B9, (d) B13, (e) B17, and (f) B21

**Width Distribution of Gold Nanorods on the basis of FESEM Image**

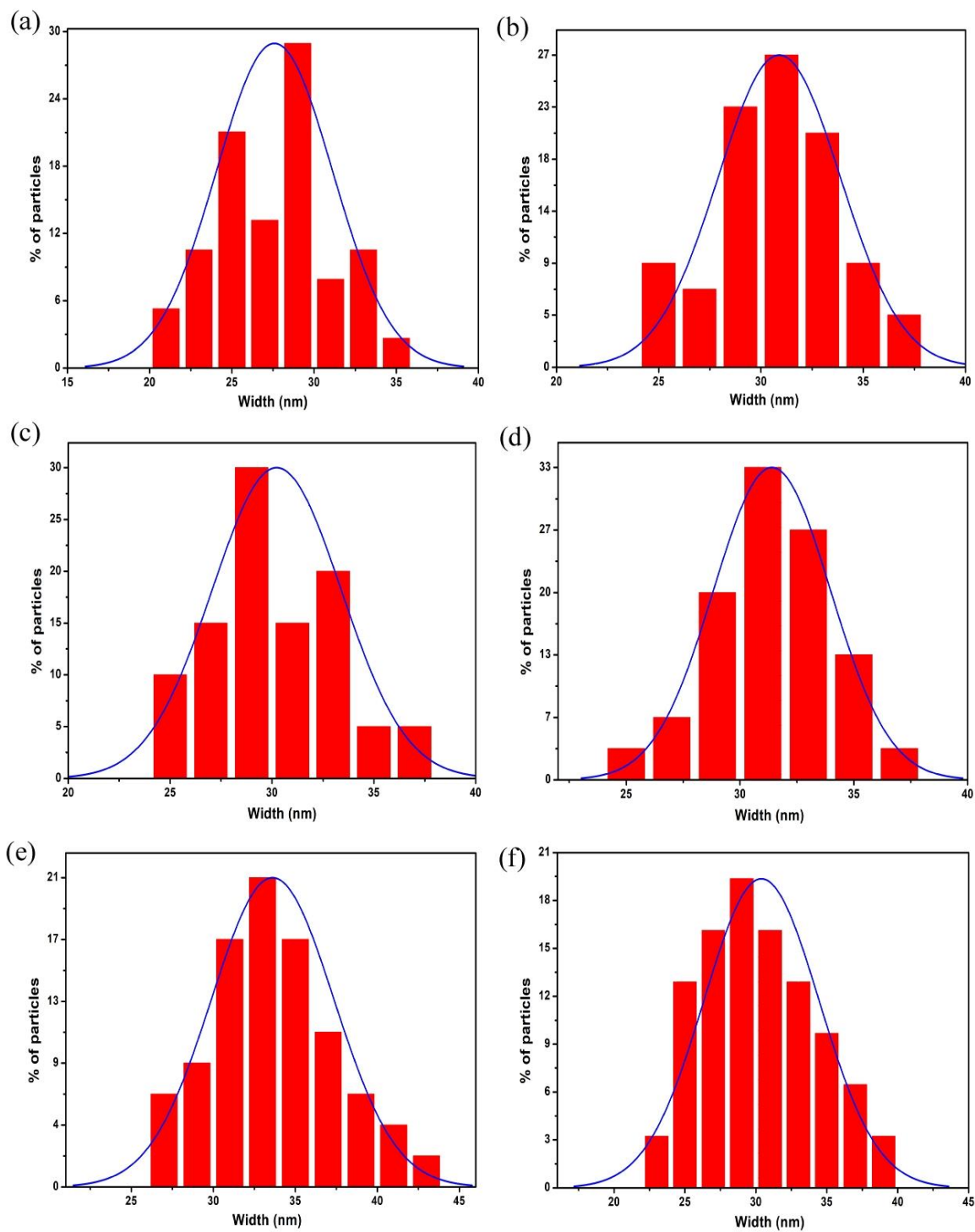

**Figure S6.** The histogram shows the width distribution of gold nanorods. The average particle size has been calculated from the Gaussian distribution function fitting as shown by the solid blue curve for the FESEM images (Fig. 6 of manuscript) of sample (a) B1, (b) B5, (c) B9, (d) B13, (e) B17, and (f) B21

**Length Distribution of Gold Nanorods on the basis of TEM Image**

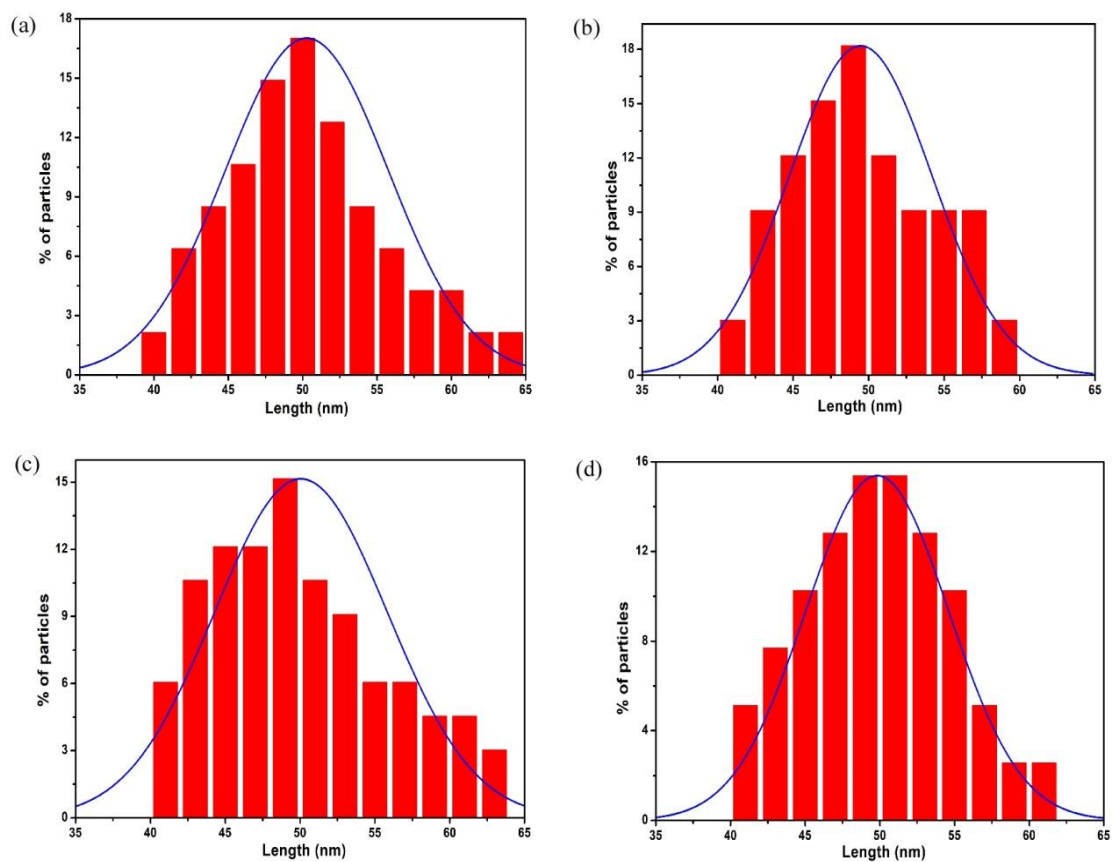

**Figure S7**. The histogram shows the length distribution of gold nanorods. The average particle size has been calculated from the Gaussian distribution function fitting as shown by the solid blue curve for the TEM images (Fig. 7 of manuscript) of sample (a) B1, (b) B5, (c) B9, and (d) B13.

**Width Distribution of Gold Nanorods on the basis of TEM Image**

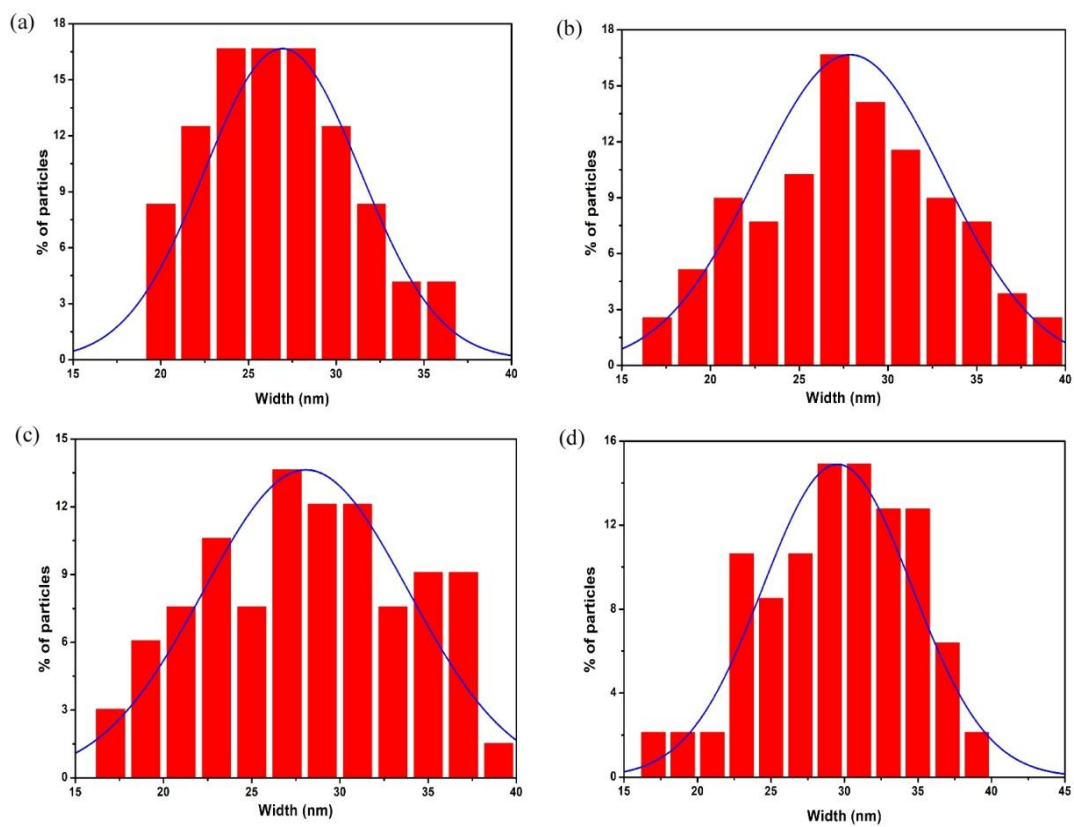

**Figure S8.** The histogram shows the width distribution of gold nanorods. The average particle size has been calculated from the Gaussian distribution function fitting as shown by the solid blue curve for the TEM images (Fig. 7 of manuscript) of sample (a) B1, (b) B5, (c) B9, and (d) B13.

**EDS Analysis of Gold Nanorods**

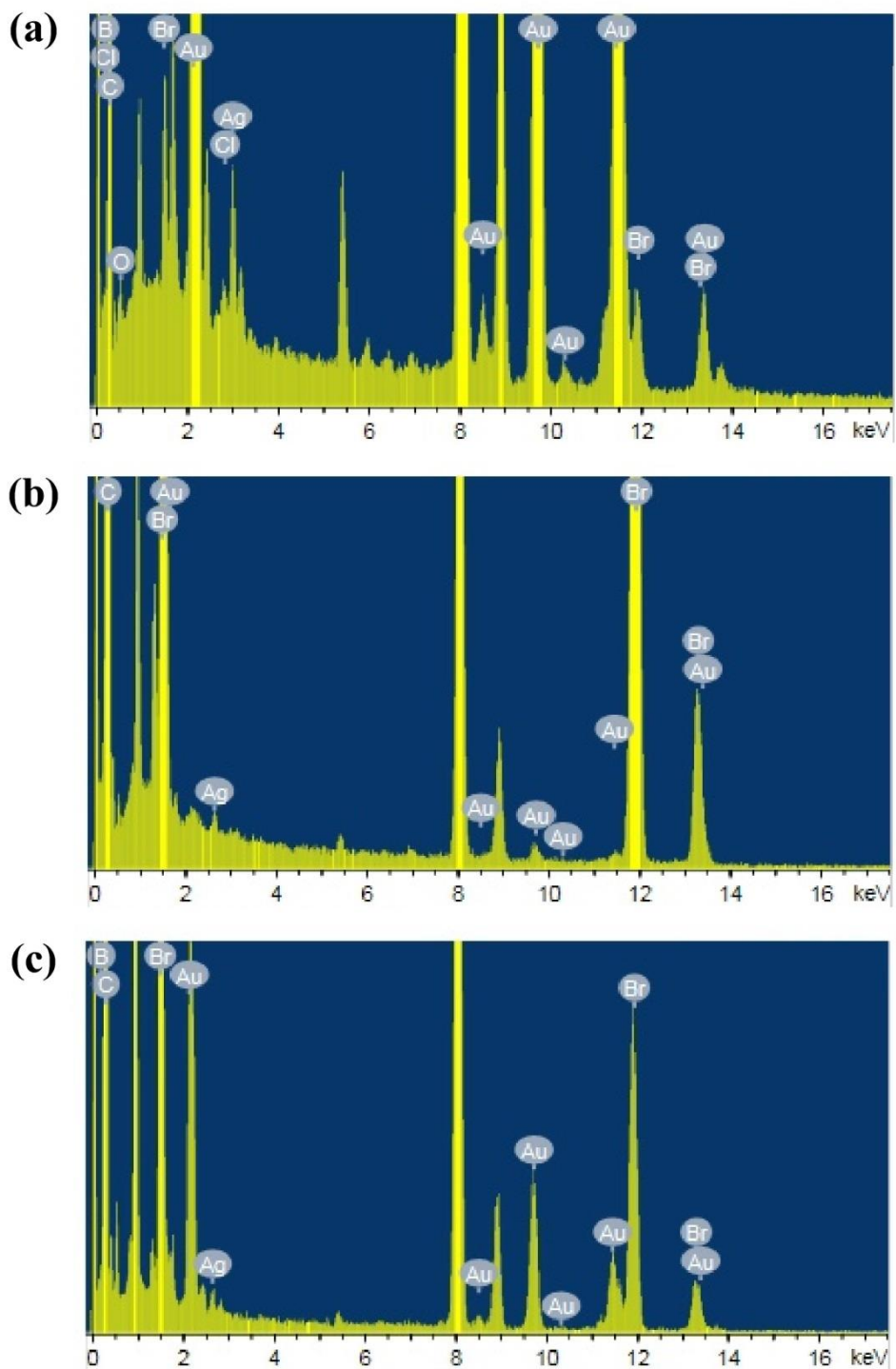

**Figure S9.** Energy dispersive x-ray spectroscopy (EDS) analysis corresponding to (a) B1, (b) B5, and (c) B9 samples (see Table - S3). Each spectrum in EDS shows the presence of Au along with other elements used in the synthesis of Au nanorods.